\documentstyle[aps,epsfig]{revtex}
\newcommand{\be}{\begin{equation}}
\newcommand{\ee}{\end{equation}}
\newcommand{\beq}{\begin{eqnarray}}
\newcommand{\eeq}{\end{eqnarray}}

\newcommand \la{\raisebox{-.5ex}{$\stackrel{<}{\sim}$}}

\begin{document}
{\Large
\begin{center}
\bf{Coulomb Effects on Particle  Spectra 
in Relativistic Nuclear Collisions}
\end{center} }

\begin{center}
H.W. Barz$^1$, J.P. Bondorf$^2$, J.J. Gaardh{\o}je$^2$, 
and H. Heiselberg$^3$\\[5mm]
$^1$ Forschungszentrum Rossendorf, 
Pf 510119, 01314 Dresden, Germany\\
$^2$Niels Bohr Institute, Blegdamsvej 17, 2100 Copenhagen, Denmark\\ 
$^3$NORDITA, Blegdamsvej 17, 2100 Copenhagen, Denmark

\vspace*{10mm}
\today\\
\end{center}
\vspace*{10mm}
{\bf Abstract:}\\
Coulomb effects on $\pi^\pm$ and $K^\pm$ spectra in
relativistic nuclear collisions are  investigated.  
At collision energies around 1 GeV the ratio of $\pi^-$ to $\pi^+$
is enhanced several times at low transverse momenta
but less at ultrarelativistic
energies. We describe the ratios at SIS, AGS and SPS energies
with simple analytic models as well as more elaborate numerical
models incorporating the expansion dynamics.
The Coulomb effect depends on the properties of the source after 
the violent collision phase and
provides information on source sizes, freeze-out times, and 
expansion velocities.  Comparison with results from HBT analyses
are made. Predictions for $\pi^\pm$ and $K^\pm$ at RHIC and LHC energies 
are given.\\[10mm]


\vspace*{10mm}


\section{Introduction}

An asymmetry in the number of opposite
charge pions has been observed  
at intermediate energies \cite{early} and has recently
also been identified in heavy ion collisions at 
energies around 1 $A\cdot$GeV \cite{Wagner,Pelte}, 11.4 $A\cdot$GeV
\cite{AGS}, and 158 $A\cdot$GeV \cite{NA44}.
The effect is strongest at lower beam energies. 
The ratio of negative
to positive pions at low pion momenta
is $\sim 3$ at SIS energies, but only $\sim 1.6$ at AGS and SPS energies
for the central collisions of heavy nuclei as $Au$ or $Pb$.
A possible cause for this effect is the Coulomb interaction between 
the produced pions and the positive charge from the reaction 
partners.

Earlier work \cite{Gyulassy,Stock,Bao,Osada} has described the $\pi^-/\pi^+$
enhancement/reduction by Coulomb attraction/repulsion from a static source. 
In this paper we underline
the importance of the source dynamics and
of the longitudinal and transverse expansion of the fireball. 
We emphasize the different reaction kinematics and
dynamics that evolve from 1 $A\cdot$GeV to 160 $A\cdot$GeV.
At the lower collision energies nuclear matter is fully stopped and expands
relatively slowly in all directions whereas at higher energies
much of the initial motion survives 
and the system expands faster preferentially in longitudinal direction.

Since the Coulomb forces influence the matter essentially after
freeze-out, the asymmetry in the number of differently charged
particles can be directly related to the freeze-out parameters of the
matter.  Therefore, a detailed description of the violent initial part
of the collision is unnecessary, and it is sufficient to consider the
state of the nuclear matter at the time $\tau_f$ of freeze-out.
Similar to particle interferometry measurements the observation of
Coulomb effects can help to determine the dynamics of heavy ion
collisions.

We will develop approximate models by assuming simple geometrical
shapes of the charge distribution at freeze-out.
In these models we can estimate analytically the
ratio of $\pi^-$ to $\pi^+$ as a function of the transverse momentum.
In an earlier publication \cite{Barz} we have presented
a model of an expanding pion gas created in heavy ion
collisions at ultrarelativistic energies. This model is well 
applicable in a wide range of collision energies. 
We will use it
for calculating particle ratios as a check of the simpler models.

The manuscript is organized as follows. In section II we study slowly
expanding and spherically symmetric sources from which pions escape
rapidly. Such a scenario may be applicable to heavy ion collisions
around 1 $A\cdot$GeV and we compare to SIS data.  In section III, we
study rapidly expanding systems from nuclear collisions at
ultrarelativistic energies, which may apply to collisions at the AGS
and SPS. We compare to recent AGS and SPS data. In section IV we
investigate the situation for particles at very low momentum. In
section V we briefly overview the main ingredients of the dynamical
model developed in ref. \cite{Barz} and compare.  In section VI we
compare freeze-out radii to those obtained from HBT analyses.  In
section VII we discuss the influence of particles created at late
stages from resonance decays, and in section VIII we consider  $K^-/K^+$
ratios. Finally, predictions for charged pion and kaon ratios at RHIC
and LHC energies are given.

\section{Slowly Expanding Charge}

In nuclear collisions charged particles feel a Coulomb force due
to the net positive charge originating form the protons in the colliding
nuclei. In particular the light produced particles as $\pi^\pm,K^\pm,...$ 
moving slowly with respect to the net positive charge are significantly
affected by the Coulomb field.
We will concentrate on pions, which are affected strongest by the 
Coulomb field since they are the lightest particles, but all results apply to
any particle of mass $m$ as, e.g., kaons or protons.

\begin{figure}
\centerline{
\psfig{figure=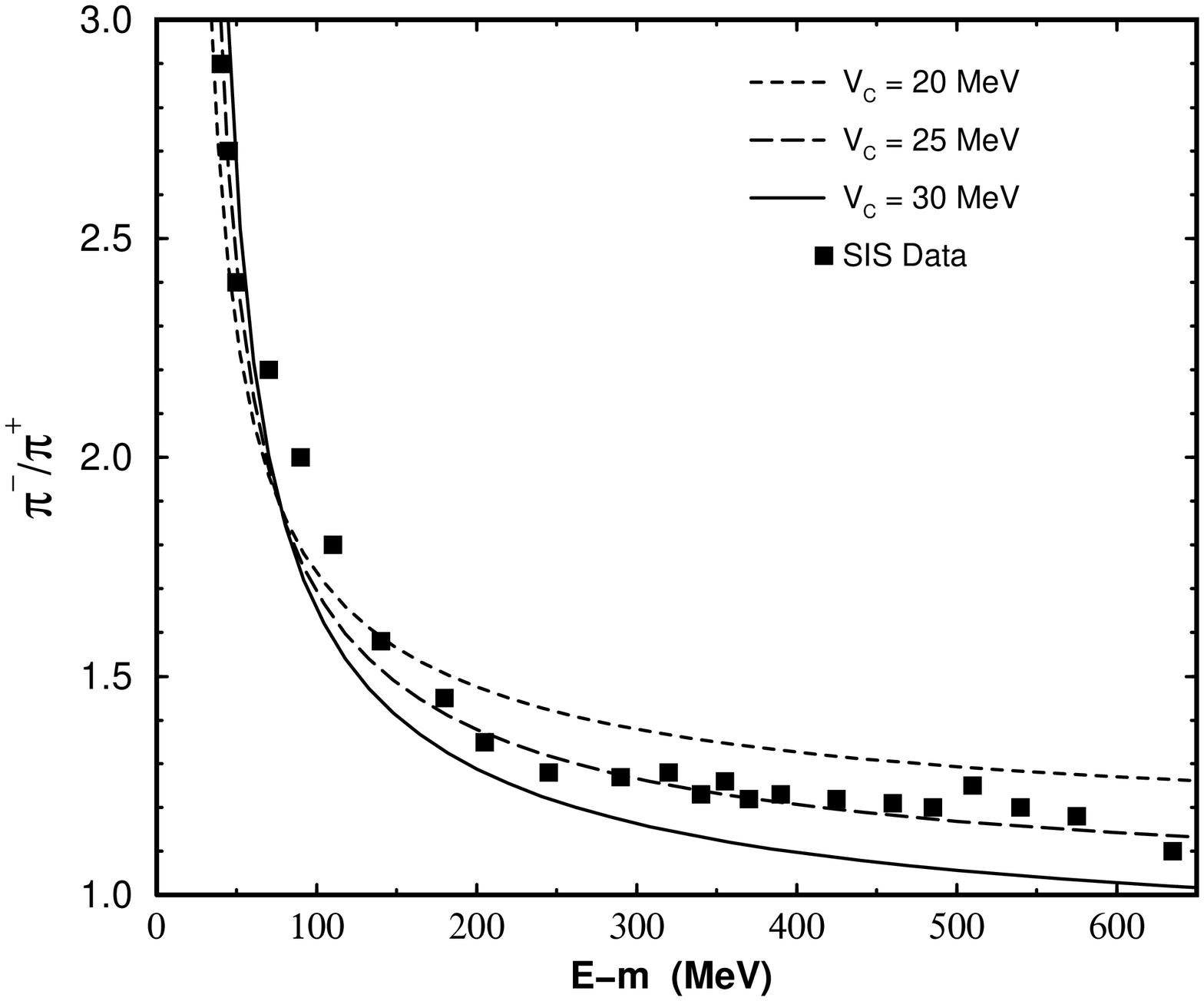,width=10cm,height=15cm,angle=-90}}
\caption{Experimental $\pi^-/\pi^+$ ratios
\protect\cite{Wagner}
for the reactions $Au+Au$ at 1.0 AGeV
as a function of the pion kinetic energy
around 44 degrees from forward direction in the lab system 
(corresponding to ca. 90 degrees in cms).
Curves show calculations 
according to Eq. (\ref{ratio}) for the case of
a slow expanding spherical source using various
Coulomb energies $V_c$.
Using a slope parameter of 75 MeV
a value of $V_c\simeq 27$ MeV
gives good agreement with data (full squares)\protect\cite{Wagner}.
\label{sisfig}  }
\end{figure}

The net charge of the source resides mainly in protons also after the
collisions. Expansion of the ``cloud of charge'' can thus be considered slow
when typical proton velocities $v_p$ are much smaller than pion velocities
$v_\pi$. 
At SIS energies the energies of protons from central collisions and
at midrapidity follow 
approximately a Boltzmann distribution with temperature
$T_p \simeq 100 $ MeV \cite{Muentz}. 
Only slow pions with  $v_\pi < v_p$ are affected while for pions with
kinetic energies\footnote{We will in the following use the convention
$\hbar=c=k_B=1$.} 
\beq
  \frac{1}{2}m_\pi \langle v_\pi^2\rangle \; > \;
  \frac{1}{2}m_\pi \langle v_p^2\rangle 
  \simeq \frac{3}{2}T_p\frac{m_\pi}{m_p}  \label{EminSIS}
\eeq
the expansion can be ignored.
At SIS energies this limit is thus $\sim20$ MeV and Eq.
(\ref{EminSIS}) applies to all SIS data 
shown in Fig. \ref{sisfig}. Thus the charge expansion can be ignored for
the SIS data.

Let us therefore consider a particle ($\pi^\pm,K^\pm,...$)
with mass $m$ and energy $E_0=\sqrt{m^2+p_0^2}$ which is
formed in a spherically symmetric
source with net charge $Z$ evenly distributed over the volume of radius $R$.
We will for the moment assume that the charge is static and will later
extend the analysis to an expanding charge.
The Coulomb potential varies from $Ze^2/R$ at the surface to 3/2 times that
value at the center with the average
\beq
   V_c=\frac{6}{5} \frac{Ze^2}{R} \,. \label{Vc}
\eeq
The energy and momenta of $\pi^0$'s are unaffected by the Coulomb field
and define the reference  values, $E_0$ and $p_0$, whereas
the $\pi^\pm$ are accelerated/decelerated in the field.
As the volume expands slowly with respect to the fast pions,
the final $\pi^\pm$ energy is 
\beq
  E = E_0 \pm V_c \, . \label{Epi}
\eeq

The effect on the 
particle distribution functions can be derived from
particle conservation, i.e., the number of particles before Coulomb
effects are not changed after, $dN_0$=$dN$. The $\pi^{\pm}$ pion distributions
at freeze-out $d^3N_0/d^3p_0$ now have to be
transformed by changing their energy and momentum 
following (\ref{Epi}). The final distribution is
\beq
   \frac{dN}{d^3p} = \frac{dN_0}{d^3p_0} \frac{d^3p_0}{d^3p}
   = \frac{dN_0}{d^3p_0} \frac{p_0}{p} \frac{E_0}{E} \,.
\label{dNd3p}
\eeq
The original $dN^0/d^3p_0$ can be taken from $\pi^0$
data around 1 GeV. These pion spectra are
to a good approximation given by 
$dN^0/d^3p_0 \sim\exp(-E_0/T)$ with slope parameter $T$. 
With final $\pi^\pm$ energy and momentum given by (\ref{Epi}) and 
using (\ref{dNd3p}) for
both $\pi^-$ and $\pi^+$, one obtains the ratio (see also \cite{Bao,BB})
\beq
  \frac{\pi^-}{\pi^+} \equiv \frac{dN^-/d^3p}{dN^+/d^3p} \quad
  = \langle \frac{\pi^-}{\pi^+}\rangle
 \exp\left(-\frac{2V_c}{T}\right) \, \frac{E+V_c}{E-V_c}
  \, \sqrt{\frac{(E+V_c)^2-m^2}{(E-V_c)^2-m^2}} \, . \label{ratio}
\eeq
The square root denotes the ratio of the initial 
$\pi^-$ and $\pi^+$ momenta when they are detected with final
energy $E$. It leads to diverging rate as the kinetic energy $(E-m)$
approaches $V_c$, $\pi^-/\pi^+\propto(E-m-V_c)^{-1/2}$. 
The average $\langle \pi^-/\pi^+\rangle$ is the ratio of 
the total number of negative and  positive pions produced.
At collisions energies of 1 GeV 
most pions are produced through the $\Delta$ resonance,
the ratio is \cite{Wagner,Stock}
\beq
  \langle \frac{\pi^-}{\pi^+}\rangle=\frac{5N^2+NZ}{5Z^2+NZ}
  \simeq 1.94 \,, \label{NZ}
\eeq
where $N$ and $Z$ are the neutron and proton numbers of the $Au+Au$
nuclei before collision.

The simple formula (\ref{ratio})
gives a good description of the $\pi^+/\pi^-$
ratio for the 1 GeV $Au+Au$ collisions at SIS and allows us to
determine both $T\simeq 75$ MeV and $V_c\simeq 27\pm2$ MeV, see
Fig. \ref{sisfig}. 
These two parameters can also be estimated independently.
The measured single particle slope $T\simeq75$ MeV \cite{Muentz}
agrees nicely with our estimate.
According to Eq. (\ref{Vc}) the value of $V_c\simeq 27$MeV
and $Z=2\times 79 \times 0.75 $ (14 \% centrality) corresponds to
an average source size of $R\simeq 8.0$ fm, which is larger than
the geometrical radius of the initial participant zone of the
two colliding $Au$ nuclei. 

The corresponding baryon density at pion freeze-out is
$\sim 2A/(4\pi R^3/3)\simeq 0.85 \rho_0$, i.e.,
somewhat smaller than normal nuclear matter 
density $\rho_0$.
We will see that this result agrees with the investigations in
the dynamical model described in section V.
In the analysis of this data within the transport model \cite{Teis}
a similar value of the Coulomb energy was obtained, although at
smaller density of $\sim0.6\rho_0$.

\section{Rapidly Longitudinally Expanding Systems}

At the higher AGS, SPS, RHIC and LHC energies the systems expands rapidly in
particular in the longitudinal direction along the beam axis.  For
systems expanding more rapidly than the pion velocity the net charge is
reduced with increasing time resulting in decreased Coulomb effects.
At ultrarelativistic speeds one also needs to take retardation
effects of the Coulomb fields into account.

\begin{figure}
\centerline{
\psfig{figure=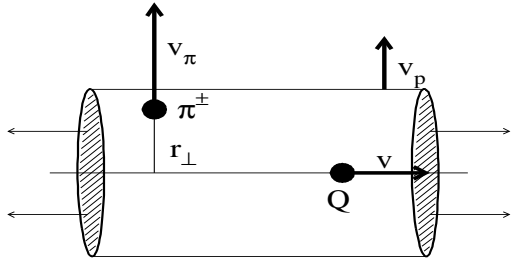,width=10cm,height=8cm,angle=0}}
\caption{
Sketch of the ultrarelativistic collision where the pion
at the  transverse distance $ {\bf r}_\perp$ 
is influenced by a charge $Q$ moving with velocity $v$ in
the expanding matter.
\label{Collfig} }
\end{figure}

Our starting point is the retarded time component of the 
electromagnetic potential
from a charge $Q$ moving with velocity $v$ along the beam axis
\cite{Jackson}
\beq
   \phi(r_\perp,t) = A^0 = 
      \frac{Q}{\sqrt{v^2t^2+(1-v^2)r_\perp^2}} ,\label{phi}
\eeq
where $r_\perp$ is the distance from the beam axis and the time is chosen such
that $t=0$ when the charge is closest to the point 
$r_\perp$ (see Fig. \ref{Collfig}).
The vector part of the potential points along ${\bf v}$ and is 
therefore longitudinal
and does not contribute to the transverse electric field that
affects transverse momenta. The longitudinal component even vanishes 
for an approximately constant rapidity distribution since
the contributions of negative and positive velocities cancel each other.

 In ultrarelativistic heavy ion collisions the charged particles have
large longitudinal but small transverse momenta. The distribution of
longitudinal velocities $v$ can be obtained from the measured rapidity
distributions since $v=\tanh(y)$
\beq
   \frac{dN^{ch}}{dv} = \frac{1}{1-v^2} \frac{dN^{ch}}{dy} \,.
\eeq
In central $Pb+Pb$ collisions at SPS energies the net proton
rapidity distribution
is rather constant at midrapidities, $dN^{(p)}/dy=37\pm2$ 
\cite{NA49,NA44prot}.

 Assuming an approximately flat charge rapidity distribution,
the electric field becomes
\beq
   {\bf E}(r_\perp,t) &=& -\nabla\phi(r_\perp,t) 
    = e\int  \frac{dN^{ch}}{dy} 
 \frac{{\bf r}_\perp dv}{(r_\perp^2+v^2(t^2-r_\perp^2))^{3/2}}
   \nonumber\\
   &=& e\frac{dN^{ch}}{dy} \frac{2}{t}
   \frac{{\bf r}_\perp}{r_\perp^2}
      \,. \label{E}
\eeq
In relativistic heavy ion collisions
the upper and lower velocities range from 
$\pm c$ .\footnote{The exact
expression is obtained by replacing the time $t$ in (\ref{E})
by $\sqrt{v_0^2t^2+r_\perp^2(1-v_0^2)}$, where $v_0$ is the lower or upper
velocity. Here,
$v_0<c$ limits the longitudinal extent of the charge cylinder and thus 
removes the $t=0$ singularity for nonvanishing ${\bf r}_\perp$.
Since the times and distances relevant for Coulomb effects are
$t\simeq R/c$ and  $r_\perp \sim R$ respectively,
 where $R$ is the size of the nuclei,
we can safely approximate $v_0\simeq c=1$.}
In the Bjorken scaling picture \cite{Bjorken} $t$ counts time from the moment 
$t=0$ of the total overlap of the collision partners. 

As can be seen from (\ref{phi}) the electric potential is dominated by
the slow charges $(v\sim0)$ simply because they provide a Coulomb
field for a longer time.  Thus pions with rapidity $y$ mainly feel 
the Coulomb field from charged particles at similar rapidities. If one
instead assumes a charge particle rapidity distribution peaked at
target and projectile rapidities, (i.e. the initial charge
distributions of the colliding nuclei) then one finds that the
electric potential is suppressed by a factor $\gamma_{cms}^2\sim 100$
near midrapidities at SPS energies as compared to the constant
rapidity distribution. In contrast, assuming a static central
charge the push given to the positively
charged pions is greatly overestimated.

As above we assume that the pions are trapped in the hot matter until 
freeze-out at time $\tau_f$. After this time the pions can propagate freely
only influenced by the Coulomb field. During the collision the
particle rapidity distributions change due to interactions. However,
at late times the retarded potential (\ref{phi}) arises mainly from the
free streaming phase or the flowing phase in the period between the
collision and freeze-out. 
According to Gauss law the electric field in the interior 
is reduced by an area factor $r_\perp^2/R_f^2$, where $R_f$ now is the
transverse size of the system at freeze-out, i.e.,
\beq
  {\bf E}({\bf r}_\perp,t) &=& e\frac{dN^{ch}}{dy}\frac{2}{t}
         \, \times \left\{ \begin{array}{ll}
         \frac{{\bf r}_\perp}{R_f^2} \,, &  r_\perp < R_f \\
         \frac{{\bf r}_\perp}{r^2_\perp}        
    \,, &  r_\perp >R_f  \end{array}\right\} . 
\label{Eb}
\eeq
  
A charged pion of transverse momentum $p_{\perp,0}$
at freeze-out and final momentum $p_\perp$ receives a momentum change
or ``Coulomb kick''
\beq
  \Delta {\bf p}_\perp &=& {\bf p}_\perp-{\bf p}_{\perp,0}
  = \pm e \int^{\infty }_{\tau_f} {\bf E}(r_\perp,t) dt
     \,, \label{Dpi} 
\eeq
where the $\pm$ refers to positive and negative pions, respectively.
If the Coulomb kick is small so that the pion velocity changes little,
the position of a $\pi^\pm$ at freeze-out ${\bf r}_f$ with  
velocity ${\bf v}_\pi$ developes according to
\beq
   {\bf r} = {\bf r}_f +{\bf v}_\pi (t-\tau_f) \,. \label{trajec}
\eeq
For an ensemble of
pions we obtain the momentum change by
averaging over the freeze-out position ${\bf r_f}$ 
in Eq. (\ref{Dpi}). The result will generally
depend on both $R_f$ and $\tau_f v_\perp$. We find, however, that 
the momentum change is only a slowly decreasing function
of the freeze-out time. For a freeze-out time of 
$\tau_f=8$fm/c, $R_f\sim10$fm and with pion velocities as given by the
low $m_\perp$ data, we obtain the estimate
\beq
  p_c \equiv |\Delta p_\perp| &\simeq& 
   2e^2 \frac{dN^{ch}}{dy} \frac{1}{R_f} \, 
     \,, \label{Dp} 
\eeq
in the direction of $\pm{\bf p}_\perp$.
The variation with freeze-out time is $\pm$10\%
for $v_\perp\tau_f/R_f$ in the range $0.5-0.8$.

It is important to note that the difference between results for the
static charge of Eqs. (\ref{Vc},\ref{Epi}) and the rapidly expanding
charge of Eqs. (\ref{Dpi},\ref{Dp}). In the former case the field
decreases at large distances as $r^{-2}$ whereas in the latter it
decreases as $t^{-1}r_\perp^{-1}$. As a consequence, the {\it energy}
is shifted by an amount proportional to the charge $Z$ in the former
case while in the latter the {\it momentum} is shifted by an amount
proportional to $dN^{ch}/dy$. In both cases, however, the scale of the shifts
is determined by the source size $R_f$ at freeze-out.

As above we can derive the Coulomb effect on the transverse
particle distribution functions 
\beq
   \frac{dN}{d^2p_\perp} = \frac{dN_0}{d^2p_{0,\perp}}\, 
  \frac{p_{0,\perp}}{p_\perp}  \,.
\label{dNd2p}
\eeq
The pion spectra are well reproduced by
$dN/d^2p_\perp\propto \exp(-m_\perp/T)$ with
$T\simeq150$ MeV at both AGS and SPS energies \cite{NA44slopes}. 
Defining $m^\pm_\perp =\sqrt{m^2+(p_\perp\pm p_c)^2}$ and using
Eqs. (\ref{Dpi}) and (\ref{Dp}) 
\beq
   \frac{\pi^-}{\pi^+} \equiv \frac{dN^-/d^2p_\perp}{dN^+/d^2p_\perp}
   = \langle\frac{\pi^-}{\pi^+}\rangle
     \exp\left(\frac{m^-_\perp-m^+_\perp}{T}\right)
    \frac{p_\perp+p_c}{p_\perp-p_c} \,.  \label{ratio3}
\eeq

\begin{figure}
\centerline{
\psfig{figure=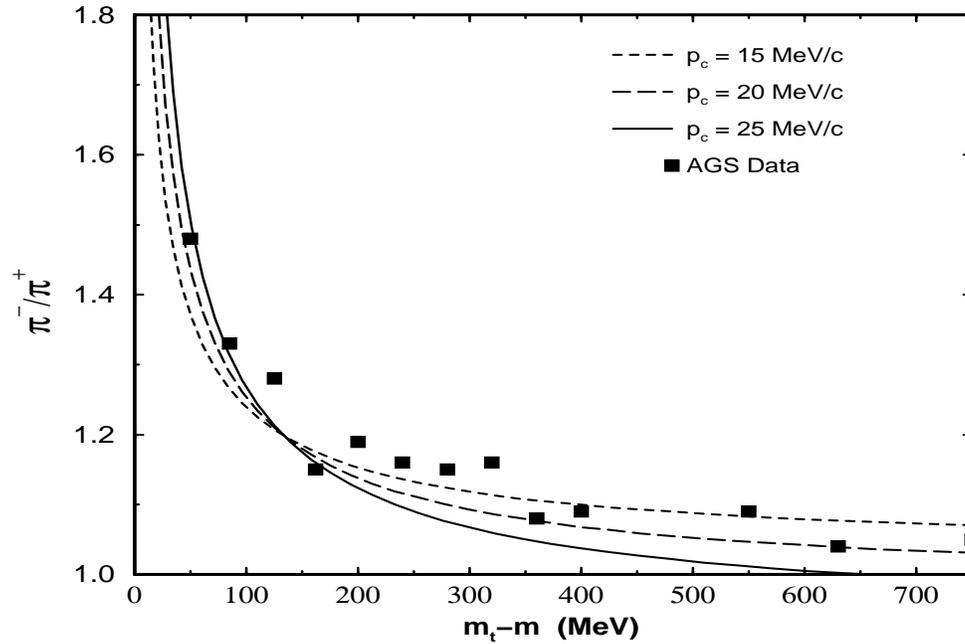,width=10cm,height=15cm,angle=-90}}
\caption{
 Experimental $\pi^-/\pi^+$ ratios \protect\cite{AGS}
for the reactions $Au+Au$ at 11.6 AGeV
as a function of transverse mass compared to
calculations employing Eq. (\ref{Dp}) using a constant momentum shift
$p_c$. 
\label{agsfig} }
\end{figure}

The pion ratio predicted by (\ref{ratio3}) is compared to AGS and SPS
data in Figs. \ref{agsfig} and \ref{spsfig}. Here the experimental
values $\langle \pi^-/\pi^+\rangle\simeq 1.2$ and $1.05$ at AGS
\cite{AGSslopes} and SPS \cite{SPSratio} energies, respectively, have
been used.  These values are closer to unity than at SIS energies due
to the abundant pion production not connected to
$\Delta$-resonances. The experimental data is well reproduced with
$\Delta p_\perp^{AGS}\simeq 20$ MeV/c and $\Delta p_\perp^{SPS}\simeq
12$ MeV/c.  From the measured proton rapidity distribution,
$dN^p_{AGS}/dy\simeq 70$ and $dN^p_{SPS}/dy\simeq 37$, and
Eq.(\ref{Dp}) we can now extract the size of the systems at
freeze-out: $R_f^{AGS}\simeq 10$ fm and $R_f^{SPS}\simeq 9$ fm. The sizes
are substantially larger than the geometrical radii $R_{geom}\simeq
6\sim R_{Pb}$ fm of the collision zones (for 14\% centrality)
indicating that significant expansion takes place before freeze-out.

For the $S+S$ collision at the SPS $dN^p/dy\simeq 4$. Assuming
$R_f\simeq 4$ fm we find a rather small value for $p_c\simeq 4$MeV/c, and
the corresponding Coulomb effect is thus very small as seen in
Fig. \ref{spsfig2}.

\section{Slow Particles}

Slow pions, i.e. pions
with velocities $v_\pi$ smaller than the expansion velocity 
of the net charge which is typically given by the proton velocities $v_p$,
are less affected by the Coulomb field.
According to Eq. (\ref{EminSIS}) such slow pions should have kinetic 
energies below 20 MeV at SIS energies and transverse kinetic energies
below 40 MeV at SPS energies.

We can estimate the effect of expansion by simply assuming
that the protons solely determine the expansion of the charge.
 If we assume that the proton velocities are distributed according
to a nonrelativistic Boltzmann distribution with temperature $T_p$,
the charge should therefore be reduced by a factor
\beq
   {\cal R}_n^{exp} &=& 
   \frac{\int_0^{v_\pi}\exp(-\frac{1}{2}m_pv^2/T_p)\; v^{n-1}dv}
        {\int_0^\infty \exp(-\frac{1}{2}m_pv^2/T_p)\; v^{n-1}dv} 
   \; =\; 
    \left\{ \begin{array}{ll}
          1\,-\, \exp(-x^2),  \,  & n=2 \\
          erf(x)-(2/\sqrt{\pi})x\,\exp(-x^2), \, & n=3
    \end{array}\right\}   \, ,
  \label{Reduc}
\eeq
where $x=\sqrt{\frac{1}{2}m_pv_\pi^2/T_p}$ and $erf(x)$ is the error
function.
The exponent $n=2$ is for cylindrical geometry with transverse
expansion and $n=3$ is for spherical expansion. 
Reducing the charge (or proton rapidity density) by this factor
reduces the Coulomb kick $V_c$ (or $p_c$) is diminished by 
an amount which now depends on the particle energy. This fact needs
to be properly included in the Jacobian $dp_{0,\perp}/dp_\perp$ 
of the resulting charge particle distribution (Eq. (\ref{dNd2p}))
The result of transverse expansion is shown in Fig. \ref{rspsfig}
for $\pi^-/\pi^+$ with $p_c=15$ MeV/c and in Fig. \ref{kspsfig} for
$K^-/K^+$. 

The factor (\ref{Reduc}) overcorrects for transverse expansion since
it completely removes the Coulomb corrections for $v_\pi=0$.
However, even for $v_\pi=0$ the pions are still accelerated
by the Coulomb field which decreases as $t^{-1}$.
We therefore consider slow pions embedded in a charge expanding with
velocity $v_p$.
Pions  with velocities smaller than $v_p$ stay inside
the expanding source in the potential decreasing with time.
As the net positive charge $Z$ or $2dN^p/dy$ expands spherically or
transversely with velocity $v_p$,
the radius increases with time
\beq
   R = R_f +v_p (t-\tau_f) \,. \label{R}
\eeq
Here, $R_f$ is the radius at freeze-out radius at freeze-out time $\tau_f$.
The pion moves according to Eq. (\ref{trajec}) and we will find 
that the Coulomb effect leads only to minor
changes in pion momenta, hence  to leading order 
we can assume that ${\bf v}_\pi$ is constant.

In the case where pions move slowly in the electric field 
\begin{equation}
{\bf E}(t) =  Ze \frac{{\bf r}}{R^3} \,. \label{Esphere}
\end{equation}
of a sphere  with a constant charge density
the net momentum change is 
\beq
\Delta {\bf p} &=& \pm e\int_{\tau_f}^\infty {\bf E}(t)\, dt 
      = \pm \frac{Ze^2}{2 R_fv_p}
       \left[\frac{{\bf v}_\pi}{v_p}+\frac{{\bf r}_f}{R_f} \right]
      \,,   \label{pgain}
\eeq
for positive and negative pions, respectively.
Retardation effects are ignored for simplicity
which is a rather good approximation
for spherical systems where the charge moves primarily in radial
direction (see also appendix A).
The average of the freeze-out position
vanishes, $\langle {\bf r}_f\rangle=0$, if there is no
transverse flow.
Therefore, averaging over ${\bf r}_f$ will to first approximation
only lead to a smearing of the $\pi^+$ and $\pi^-$
distributions and by the
{\it same} amount. When taking ratios this effect cancel
out, and we therefore ignore it in the following.
The final momentum is thus
\beq
  {\bf p} &=& {\bf p}_0 (1\pm C_3)\,,\quad  
   C_3 = \frac{Ze^2}{2m_\pi v_p^2 R_f} \,, \label{Dps}
\eeq
i.e., the positive and negative pion momenta are scaled up
and down respectively.

The case of rapid longitudinal expansion was studied in the previous
section. The resulting electric field is given by Eq. (\ref{Eb})
and from Eq. (\ref{pgain}) we
find the change in transverse momentum
\beq
  \Delta {\bf p}_\perp &=&{\bf p}_\perp-{\bf p}_{\perp,0}
  = \pm e\int^{\infty}_{\tau_f} {\bf E}(r_\perp,t) dt
  \simeq \pm e^2 \frac{dN^{ch}}{dy} \frac{2}{R_f} \, 
   \frac{{\bf v}_\perp}{v_p}    \,. \label{Dp2}
\eeq
The latter approximate expression holds under the condition
$v_p\tau_f\la R_f$. 
For small $p_\perp$ we can approximate $p_\perp=mv_\perp$
and thus find
\beq
   {\bf p}_\perp =  {\bf p}_{\perp,0}(1\pm C_2)\,\quad ;\quad
   C_2\simeq e^2 \frac{dN^{ch}}{dy} \frac{2}{mv_pR_f}
  \,. \label{pf}
\eeq

The Coulomb effect on particle distribution functions can be obtained
from particle conservation, i.e. $dN=dN_0$ is the same for initial and
final pions,
\beq
   \frac{dN}{d^np} = \frac{dN^0}{d^np_{0}}
  (1\pm C_n)^{-n}  \,.  \label{dND}
\eeq
The ratio of negative to positive pions is then
\beq
   \frac{\pi^-}{\pi^+} \equiv \frac{dN^-/d^np}{dN^+/d^np}
   = \langle\frac{\pi^-}{\pi^+}\rangle
     \left(\frac{1+C_n}{1-C_n}\right)^n
     \exp\left(\frac{E^--E^+}{T}\right) \,,  \label{ratio2}
\eeq
where $E^\pm=\sqrt{m^2+p^2(1\pm C_n)^2}$.
Notice that $C_n\ll 1$ is implicitly required since we approximated
the trajectory ${\bf r}(t)$ by assuming
constant velocity in the electric field. 

Introducing the kinetic energy $E_{kin}=E-m$ (for $n=3$) and
$E_{kin}=m_\perp-m$ (for $n=2$), the ratio decreases linearly as
\beq
   \frac{\pi^-}{\pi^+} 
   \simeq \langle\frac{\pi^-}{\pi^+}\rangle
    \left(1 +2nC_n(1 - \frac{E_{kin}}{T} )
    \right) \,.     \label{ratio5}
\eeq
for small kinetic energies, $E_{kin}\ll m$.
The ratio is thus enhanced by $2nC$ at small $m_\perp$ and
decreases linearly on a scale determined by $T$. 

It is interesting to compare our approximation with the exact analytic
result found by Ayala and Kapusta \cite{Ayala} treating the motion of
nonrelativistically positively charged particles in a Coulomb field from a
homogeneously charged expanding sphere.  With their parameters
($Z$=158, $R_f=8$ fm, $v_p$=0.2) we obtain $C_3=$ 0.7 for kaons.
This leads to a reduction of low momentum $K^+$ by a factor $(1+C_3)^{-3}=0.2$
similar to their value of 0.15 determined from Fig. 3a in
\cite{Ayala}.  However one should be aware that the suppression
factors behave asymptotically differently in the limit $v_p
\rightarrow 0$.

 Next we estimate the crucial parameters $C_2$ and $C_3$.  In central
$Au+Au$ collisions at SIS energies the proton 
velocity can be estimated from the
slope of proton spectra $v_p^2\simeq 3T_p/m_p\simeq 0.3$.  With
a central charge of $Z=122$ (15\% centrality) we obtain
$C_3\simeq0.3$ that according to (\ref{ratio2}) leads to a large
enhancement by a factor $\sim6$ at zero momentum. Unfortunately, at
SIS energies no measurements are available at very low pion energy.
For SPS we assume cylindrical geometry $n=2$. Consequently, going from $C_3$ to
$C_2$ we replace $Z$ by $2dN^p/dy$. With $v_p^2\simeq 2T_p/m_p\simeq
0.6$ we obtain $C_2=$0.12 and find a $\pi^-/\pi^+$ ratio (\ref{ratio2}) of
1.8 in agreement to the data.

As mentioned the derivation applies not only to pions but to all
particles, for example, kaons.  As the kaon is heavier we expect from
(\ref{pf}) that the Coulomb enhancement of the $K^-/K^+$ ratio is
smaller by a factor of $m_\pi/m_K\sim 1/3$.

Experiments at the AGS \cite{AGS} and SPS \cite{NA44slopes} indicate
that transverse flow is important in collisions with very heavy
nuclei. Transverse flow decreases the $m_\perp$ slopes observed in
$m_\perp$ spectra of pions, kaons and protons. The measured
``effective temperatures'' are thus larger than the intrinsic
temperatures. In our analysis transverse flow has several effects. The
temperature employed in the Boltzmann spectrum should not be the
effective slope parameter, $T\simeq 200$ MeV but the intrinsic
temperature, $T\simeq 140$ MeV.  Furthermore, flow leads to larger
$v_p$ and nonvanishing $\langle {\bf r}_f\rangle$ in (\ref{pgain}).
Particles with transverse momentum ${\bf p}_\perp$ from a thermal
source of temperature $T$ with transverse flow $u r_\perp/R_f$ arrive on
average from a point displaced from the center by $\langle {\bf
r}_f\rangle={\bf p}_\perp uR/2T$. Transverse flow thus adds a term to
(\ref{pgain}) for charged pions and usually reduces $C_n$.

We can summarize the discussion of 
Coulomb effects by addressing the important
scales entering. In the spherically symmetric case (n=3) the Coulomb
kick is given by $V_c\simeq \pm Ze^2/R_f$. For slow expansion the
particle kinetic energies are {\it shifted} (see Eq. (\ref{Epi})) by
this amount, i.e., the relevant dimensionless quantity is $V_c/m_\pi
v_\pi^2$ instead. For rapid expansion $v_p\gg v_\pi$ the kinetic energies are
instead {\it scaled} (see Eq. (\ref{Dps}) by an amount proportional to
$V_c/m_\pi v_p^2$. In the dimensionless quantity $v_\pi^2$ is
replaced by $v_p^2$ simply because it is the fastest particle velocity
that matters. A similar behavior is found for the cylindrical expansion
(n=2) except that in this case the relevant quantity is the {\it
momentum} instead of {\it kinetic energy} due to the rapid
longitudinal expansion and the Coulomb kick is $V_c\simeq
(e^2/R_f)dN^p/dy$. For slow expansion the momentum is shifted by $V_c$
(see Eq. (\ref{Dp}) and the corresponding dimensionless parameter is
$V_c/m_\pi v_\pi$. For an expansion faster than the pion velocity
the momenta are scaled by an
amount proportional to $V_c/m_\pi v_p$ (see Eq. (\ref{pf})).  Again
the pion velocity is replaced by the proton velocity since the fastest
particles determine the magnitude of the Coulomb effect.

\section{ Dynamical Model }

After the previous discussion of the typical properties of the Coulomb effect
in different energy regions we apply a more detailed
model \cite{Barz} which allows finer details of 
the freeze-out scenario.  Inspired by the Bjorken picture 
of approximate boost invariance we assume a cylindrical geometry and
include both longitudinal and transverse flow.
We introduce the following parameters: the width $\Delta y_{ch}$ of the
rapidity distribution of the charge (usually a Gaussian distribution
is taken), the thermal equilibrium temperature
$T$,  the mean transverse flow velocity $\langle \beta \rangle$, the
source radius $R_f$, and the freeze-out time $\tau_f$. 
The initial charge distribution at freeze-out is assumed to follow
a Boltzmann distribution which includes the different 
chemical potentials for positively and negatively charged particles.
Further details are explained in ref. \cite{Barz}. 

After freeze-out the different species of particles,
mainly nucleons, pions, and kaons, are assumed to move freely.
Their momentum distribution is given by a thermal
distribution superimposed on a hydrodynamic flow. 
The influence of the electromagnetic forces  on the 
emitted particles  will increase the initial charge asymmetry.
This influence is treated as a first order perturbation,
assuming that the electromagnetic field is generated by 
the charged particles moving with their unperturbed initial velocities.
This method is justified as long as the electromagnetic potential
is small compared to the mean kinetic energy of the particles.

To calculate  the electromagnetic potential $A^\mu$ 
at the space-time $x^\mu = (t, {\bf r})$
\beq
    A^\mu(t,{\bf r}) 
          = \int d{\bf r'} \frac{j^\mu(t-|{\bf r}-{\bf r'}|,{\bf r})}
            {|{\bf r}-{\bf r'}|}  \,, \label{A_mu}
\eeq
the knowledge of the currents $j^\mu(t,{\bf r})$ during the whole 
collision history is needed. Retardation effects 
are essential, especially at ultra-relativistic energies where the
longitudinal motion is not fully decelerated. 
To describe the situation before freeze-out
we have in ref. \cite{Barz} used a simple hydrodynamical scenario 
to describe the situation before freeze-out.  
We assume 
a linearly increasing longitudinal and transverse flow of the matter
between the overlap time $t=0$ and the freeze-out time $t=\tau_f$, i.e., 
the radius at 
freeze-out is $R_f= R_{geom} + 0.5\tau_f\beta_S$, where  $R_{geom}$ 
is the initial or geometrical radius of the overlap zone 
at the collision time and $\beta_S$ the velocity  at the surface 
of the source. The linear motion of the particles allows one
to calculate the net 
Coulomb mean field as a superposition of the electromagnetic
fields of the individual charged particles.
If the particles denoted by $i$ with charge $q_i$, mass $m_i$, 
and four-velocity $u^\mu_i = (1, {\bf v})/\sqrt{1-{\bf v}^2}$ 
start at space-time $y_i^\mu$ we get
\beq
 A^\mu(x^\mu) = \sum_{i} q_i \frac{u^\mu_i}
                 {\sqrt{[(u_i)_\nu (x^\nu - y_i^\nu)]^2 
           - (x^\nu - y_i^\nu)^2}}.  \label{A_sum}
\eeq
Since the calculations of retardation are very cumbersome we 
also discuss approximations in appendix A.

The motion of the particles is treated relativistically 
in the potential (\ref{A_sum})
\beq
m_i \frac{d}{d\tau_f} u_i^\mu= q_i \, u_i^\nu
        \left(\frac{\partial A_\nu}{\partial x_\mu}-
        \frac{\partial A^\mu}{\partial x^\nu}\right)
\; ,
\label{eq_motion}
\eeq 
where $\tau_f$ denotes the proper time of the particle.
The $\pi^-$ and $\pi^+$ spectra are finally calculated
by a sampling the final momenta over a set of solutions obtained 
with different initial conditions obtained from the freeze-out 
parametrization.  

In Figs. \ref{sisfig2}-\ref{spsfig2} we show the
calculations for the three above mentioned experiments.
We concentrate on the sensitivity to variations of the
freeze-out time.  The flow velocity and temperature were determined from
fits to the $m_\perp$ slope of available spectra for
particles with different masses, see
refs. \cite{Muentz,NA44slopes,AGSslopes}. For a systematic overview we
refer to ref. \cite{Herrmann}.

Fig. \ref{sisfig2} shows the result for SIS energies which has to be
compared to the schematic model of
Fig. \ref{sisfig}. For a temperature of 75 MeV and mean
transverse velocity of $\langle \beta \rangle$ = 0.3 corresponding to
a surface velocity of $\beta_s$ =0.5 good agreement to the data is
obtained. This gives a freeze-out radius of $R_f=6fm+
0.5\tau_f\beta_s$= 8 fm leading to a break-up density of 0.8
$\rho_0$. This value is just the same as that obtained in section II,
but larger than the value obtained in the transport model
\cite{Teis}.  The final production time in that model is rather long
$\sim25$ fm/c which is partially due to long lived $\Delta$
resonances. 

Figs. \ref{agsfig2} and \ref{spsfig2} 
address the ultra-relativistic regime. Here, the measured widths of the
particle rapidity distributions,
$\Delta y_{ch}$, is about 0.7 and 1.3 for AGS and SPS energies respectively.
A mean flow velocity of $\langle \beta \rangle$ =0.42 or $\beta_s=0.62$ 
was used together with a moderate temperature of 140 MeV and
120 MeV, respectively.
The results at AGS energies 
are relatively insensitive to the freeze-out time since 
the charge diffuses slower in the longitudinal direction than
for SPS energies. A freeze-out time of 10 fm/c seems to be
compatible with the data, while for SPS energies a time of about 7 fm/c
is required. 
We also show the relatively modest enhancement for the reaction
of $S$ on $S$ at SPS energies, where the rapidity density of
the charge is slightly depleted at midrapidity.  
For a more detailed discussion at SPS energies
we refer to ref. \cite {Barz}.

\section{Comparison to HBT analyses}

The source for pions, kaons and other particles created in high energy
nuclear collisions at freeze-out is also studied in two-particle
correlation functions. Such Hanbury-Brown and Twiss (HBT) analysis
provides complementary information on freeze-out time and source
size. Pion correlation functions are, however, affected by resonance
decays which actually produce the majority of the pions. Resonances
lead to an increase of the measured HBT radii, effective source sizes
and life-times \cite{HH,Wiedemann}. A significant fraction $f$ of the
pions stem from {\it long lived resonances} in ultrarelativistic
collisions which reduces the correlation function by a factor
$\lambda^{\pi\pi}\sim (1-f)^2$. Experimentally,
$\lambda^{\pi\pi}\simeq 0.5-0.6$ \cite{Franz} and so $f\simeq
25$\%. Also, long lived resonances reduce the Coulomb effect by a
minor amount as discussed below. Though these and other
effects as Coulomb corrections, detector cuts, etc., are important for
a detailed analysis, we will ignore them for our qualitative
comparison.

The transverse HBT radii are usually referred to as the outward and
sideward radii in the direction along and perpendicular to the
transverse momentum of the two correlated pions, respectively.  The
outward HBT radius contains a number of fluctuations as duration of
emission and resonances life-times and are furthermore affected by
opacities and moving emission surfaces \cite{fluc}.  The sideward HBT
radius $R_s$ is more directly connected to the geometric size of the
source at freeze-out. For a source of constant pion density the sharp
cut-of radius is given by $R_f=2R_s$, and with $R_s\simeq 5$ fm
\cite{Franz} in central $Pb+Pb$ collisions at 160 A$\cdot$GeV we
obtain $R_f \simeq 10$ fm.  This number is larger than the geometrical
radius of a $Pb$ nucleus indicating that expansion takes place before
freeze-out.

In a longitudinally expanding system with Bjorken scaling the
freeze-out time is related to the longitudinal HBT radius by $\tau_f=
R_l\, \sqrt{m_\perp/T}$.
From the NA44 data $R_l\simeq 5.5$ fm
\cite{Franz} we estimate $\tau_f\simeq 8$ fm/c.
For comparison RQMD calculations
predict substantially larger values for the freeze-out time,
$\tau_f$=15 fm/c \cite{Sorge}.
Assuming that the source starts out with the mean geometrical radius
$R_{geom}$ = 6fm (for the 15\% centrality) at the collision and 
develops transverse flow $\beta_s$ 
up to freeze-out, we find a radius of 
$R_f=R_{geom}+\beta_s \tau_f/2\simeq 9$ fm, which is slightly smaller than
the freeze-out radius obtain from the sideward HBT radius.
We conclude that both freeze-out times extracted from the
longitudinal and transverse HBT radii are compatible at SPS energies
with the one we find from $\pi^-/\pi^+$ data.  
The Coulomb freeze-out radius is also comparable to the HBT radii
found at AGS energies as seen from Fig. \ref{hbtfig}.
At SIS energies we extract the HBT radius from the 
average of the $\pi^+\pi^+$ and $\pi^-\pi^-$ HBT radii \cite{AGSHBT}
by multiplying by the geometrical factor,
$R_f= \sqrt{5/3}R_{rms}\simeq 15$fm. This radius is considerably larger
than the one extracted from the Coulomb effects. 

\section{Resonance Decay}

In ultrarelativistic collisions resonance decay into pions is
important. 
If pions stem from resonances with a life time larger than the
freeze-out time, their Coulomb push is reduced.  This is important for
the long lived resonances $\omega,\eta,\eta',K^0_S$, etc.. As mentioned
above they amount to roughly $f=25$\% if
the reduction of the HBT correlation function is solely caused by long
lived resonances.
If a fraction $f$ is unaffected by the Coulomb field only the
fraction $(1-f)$ of the 
distribution function in Eq. (\ref{dNd2p}) should be transformed.
The resulting $\pi^-/\pi^+$ ratio is therefore
\beq
   \frac{\pi^-}{\pi^+}
   = \langle\frac{\pi^-}{\pi^+}\rangle
    \frac{(1-f)\frac{p_\perp+p_c}{p_\perp}e^{-m^+_\perp/T} + fe^{-m_\perp/T}}
    {(1-f)\frac{p_\perp-p_c}{p_\perp}e^{-m^-_\perp/T} + fe^{-m_\perp/T}}
     \,.  \label{ratio4}
\eeq
The effect of long lived resonances are shown in Fig. \ref{rspsfig}
for $f=0.25$ and $p_c=15$ MeV/c. As they reduce the Coulomb effect, the
best fit is now obtained for a slighly larger value of $p_c=15$ MeV/c.
If the long lived resonances produce pions at different $p_\perp$
one should use a different effective $m_\perp$ slope for those in 
(\ref{ratio4}).

Resonances may decay differently into $\pi^-$ and $\pi^+$
depending on pion momenta and thus provide another mechanism for
changing pion distributions at low $p_\perp$.  This is the case for
$\Lambda$ and $\bar \Lambda$ particles.  In central $Pb+Pb$ collision
at SPS energies the number of $\Lambda$'s produced per unit rapidity
at central rapidies is $dN^{\Lambda-\bar\Lambda}/dy=19\pm7$
\cite{NA49}.  Of these 64\% decay into $\pi^-$. Compared to
$dN^{\pi^-}/dy=160\pm10$ the $\Lambda$'s thus contribute $\sim 8\%$ to
the $\pi^-$ yield.  Since the decay energy of pions amounts to only 35
MeV in the $\Lambda$ rest frame they would lead to an enhancement of the
$\pi^-/\pi^+$ ratio of $\sim1.4$ if included. The ratio would fall off
slowly as a function of the transverse momentum.  However, only at
most a few percent of those pions can enter the detector in the NA44
measurements because the $\Lambda$'s travel $\sim60$cm
before decaying due to the long life time.  Therefore they do not
contribute to the measured $\pi^-$ yield.
Both the $Pb+Pb$ data and the $S+S$ data are compatible with the
Coulomb effect but cannot be explained by resonances alone.  The same
conclusion was drawn from RQMD analyses \cite{NA44prot}.

\section{The $K^-/K^+$ Ratio}

The $K^-/K^+$ ratio provides another way to test the Coulomb effect
and obtain information on freeze-out. We estimate $\Delta p_\perp$ by
taking the best fit to the $\pi^-/\pi^+$ data, however, correcting it
for the slightly smaller HBT radii for kaons than pions. We estimate
$\Delta p_\perp\sim 15$ MeV/c. Furthermore we take the kaon
temperature of $T_K=230$ MeV from \cite{NA44slopes} and $\langle
K^+/K^-\rangle=0.68$ from \cite{Murray}. We can then predict the
$K^-/K^+$ as shown in Fig. \ref{kspsfig}.  As the kaons are heavier
than the pions their kinetic energies are less affected by the momentum
change. Also, the kaons move slower and are more affected
by transverse expansion of the net (proton) charge (see appendix A).
A smaller number of protons remains to give the kaons a Coulomb kick and
the $K^-/K^+$ ratio is closer to unity as seen in Fig. \ref{kspsfig}.

HBT analyses show that the kaon correlation function is not reduced by
long lived resonances, since $\lambda^{KK}\simeq 1$.  On the other
hand, the absorption of $K^-$ on nucleons by $K^-N\to\Lambda\pi$ is
much larger than for $K^+$. Therefore the initial ratio $K^-/K^+$ may
differ in very heavy ion collisions from light ones since the stopping
is larger.  The measured $K^-$ spectrum \cite{Kspectr} in fact shows a
decrease at low $p_\perp$ in heavy ion collisions as compared to light
ones by almost a factor of two. Thus, the resulting $K^-/K^+$ ratio
could be even an increasing function of the transverse energy. Recent
NA44 measurements do not show a significant dependence on the transverse
energy. If the enhancement of the $K^-/K^+$ ratio predicted in Fig.
\ref{kspsfig} should exceed the measured one, this would indicate new
physics.

Kaons may also experience a strong interaction with the mean field
from hadrons. Early reports on low $p_\perp$ enhancement for $K^+$ at
AGS energies - referred to as ``cold kaons'' - led to suggestions of a
strong attractive potential \cite{Koch} that dominated the repulsive
Coulomb potential.  The cold kaons have, however, not been confirmed
in later AGS experiments.  In nuclear matter the $K^+$ interaction is
repulsive \cite{Weise} but one cannot rule out that it becomes
attractive in hot and dense matter.

\section{Summary}

 Coulomb effects are very sensitive to the expansion dynamics of high
energy nuclear collisions.  At lower energies the expansion of the
charge cloud, which consists mainly of protons, is slow compared to
typical pion velocities. The pions experience the full Coulomb kick
and the $\pi^-/\pi^+$ ratio becomes large at small pion momenta as
described in Eq. (\ref{ratio}). At higher collision energies the
expansion of nucleons is faster and exceeds that of the very slow
pions. The Coulomb field, which the pions feel, vanishes rapidly
within the typical expansion time. The resulting $\pi^-/\pi^+$ ratio
is only enhanced by $\sim 60$\% at the smallest measured transverse
momentum $p_\perp$ at AGS and SPS energies and can be approximately
described by Eq. (\ref{ratio3}). At ultrarelativistic energies the
rapid longitudinal expansion and retardation of electromagnetic fields
are important and the charge rapidity density replaces the central
charge.  The Coulomb field affects small pion transverse momenta on a
scale given by the total Coulomb energy.  The Coulomb effect on kaons
is smaller than for pions due to their larger mass.

The Coulomb effect provides information on the sources produced in
relativistic heavy ion collisions at freeze-out independent of HBT
analyses and flow studies in particle spectra. 
HBT methods analyse the
source at freeze-out, and Coulomb effects occur after freeze-out.
The measured
enhancement of the $\pi^-/\pi^+$ ratio 
provides a severe constraint on the values found by
various methods.  From the investigations presented here the source
sizes and freeze-out times obtained from HBT analyses seem to be
compatible with those obtained from Coulomb effects. 
Calculations using  transport models like BUU or RQMD tend to
find  larger freeze-out times indicating the need for a consistent
description of the whole process to calculate charge asymmetry 
and HBT correlations.

\begin{figure}
\centerline{
\psfig{figure=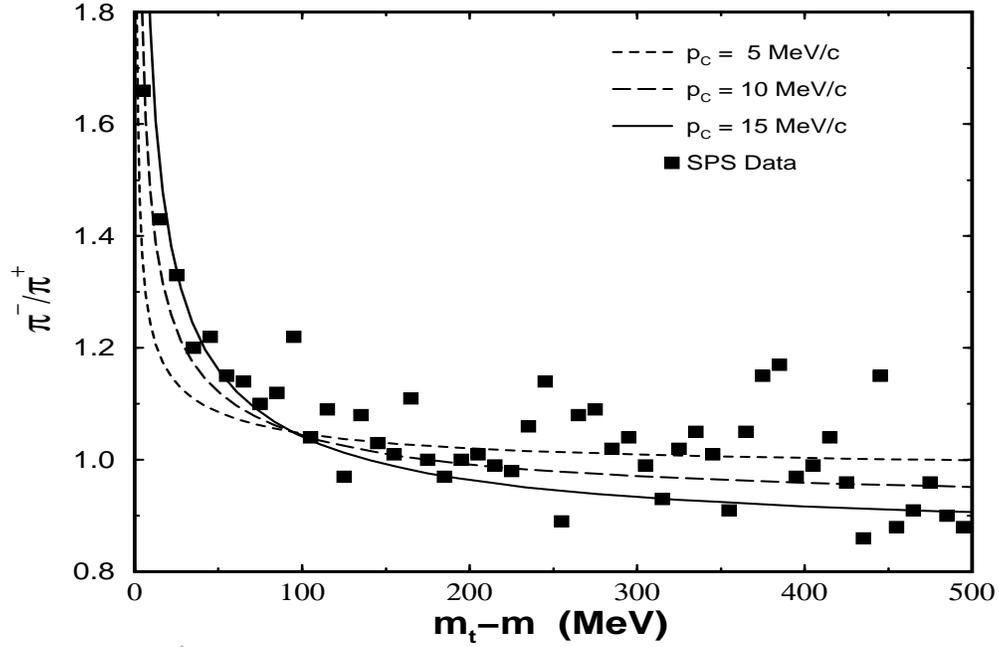,width=10cm,height=15cm,angle=-90}}
\caption{
 Experimental $\pi^-/\pi^+$ ratios \protect\cite{NA44} at SPS energies
for the reaction $Pb + Pb$ 
as a function of transverse mass compared to  
calculations using Eq. (\ref{Dp}) using a constant momentum shift
$p_c$. 
\label{spsfig} }
\end{figure}

\begin{figure}
\centerline{
\psfig{figure=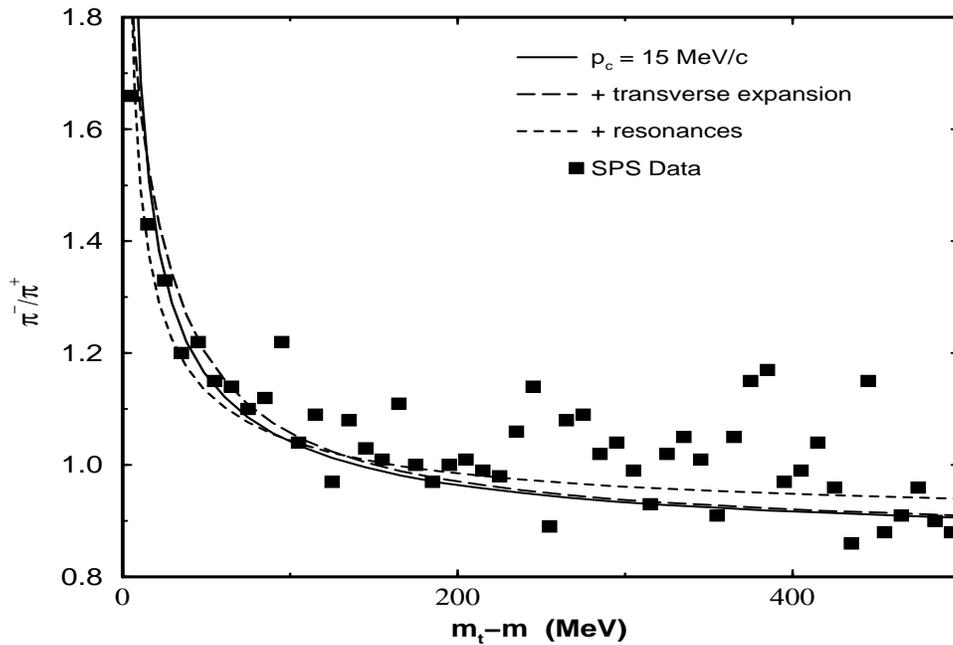,width=10cm,height=15cm,angle=-90}}
\caption{
The effect of resonances and transverse expansion on 
$\pi^-/\pi^+$ ratios Eq. (\ref{Dp})  (see Fig. \ref{spsfig} and text).
\label{rspsfig} }
\end{figure}

\begin{figure}
\centerline{
\psfig{figure=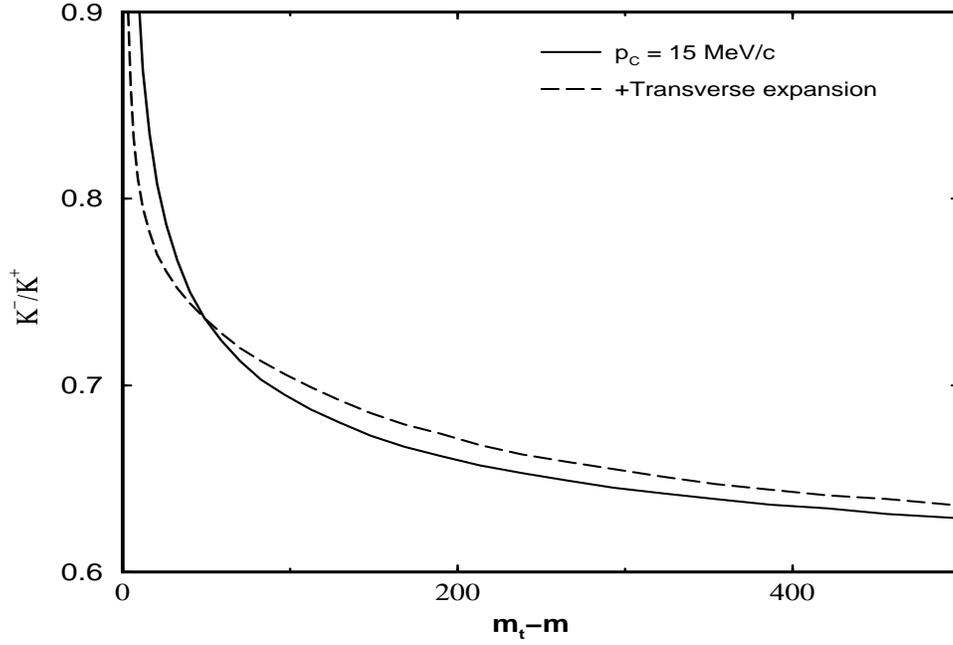,width=10cm,height=15cm,angle=-90}}
\caption{
The predicted $K^-/K^+$ ratios at SPS energies
for the reaction $Pb + Pb$ 
as a function of transverse mass from  
Eq. (\ref{Dp}) with $\Delta p_\perp=15$ MeV/c (full curve). 
Dashed curve includes transverse expansion of Eq. (\ref{Reduc}).
\label{kspsfig} }
\end{figure}

At RHIC and LHC energies smaller stopping is expected than at SPS and
AGS energies, and the charge density $dN^p/dy$ will be smaller at
midrapidity but larger near target and projectile rapidities.  Given
the amount of stopping we can try to predict the resulting
$\pi^-/\pi^+$ ratios at RHIC and LHC.  The total
$\langle\pi^-/\pi^+\rangle$ production should be very close to unity
due to the abundant pion production.  The transverse size of the
system at freeze-out can be estimated from the HBT radii assuming that
the radii continue to increase slowly with collision energy as is found
experimentally up to energies $160$A$\cdot$GeV.  By extrapolating
we estimate $R\simeq 15$ fm.  If we, for example, assume $dN^p/dy=25$
we find $p_c=5$ MeV/c. The resulting $\pi^-/\pi^+$ ratio is shown in
Fig. \ref{spsfig} (not corrected for resonances and transverse proton
expansion).  However, if a phase transition to a quark-gluon plasma
occurs, these predictions may change drastically.  The freeze-out may
occur much later so that the source is more longitudinally and
transversely expanded leading to a reduced Coulomb effect.

We acknowledge very informative discussions with A. Hansen and Nu Xu
from the NA44 collaboration. Support from The Danish Natural Science
Research Council is appreciated.

\vspace*{20mm}
\appendix{ {\bf APPENDIX A: Effects of Retardation} }\\

In ref. \cite{Teis} 
approximations were applied by using the Coulomb potential
$\phi$ as the only part of the four-potential. This  potential $\phi$ is 
calculated as being $A^0$ of (\ref{A_mu}), but neglecting 
the retardation. Thus, the equation of motion simplifies to 
\beq
m_i \frac{d}{d\tau} u^k= u^0\; q_i \frac{\partial \phi}{ \partial{x^k}},
      \;\;\;k=1,2,3.
\label{eq_approx}
\eeq

We demonstrate the deviations one obtains when applying the
approximative method (\ref{eq_approx}) where retardation and 
the spatial components are neglected.

\begin{figure}
\centerline{
\psfig{figure=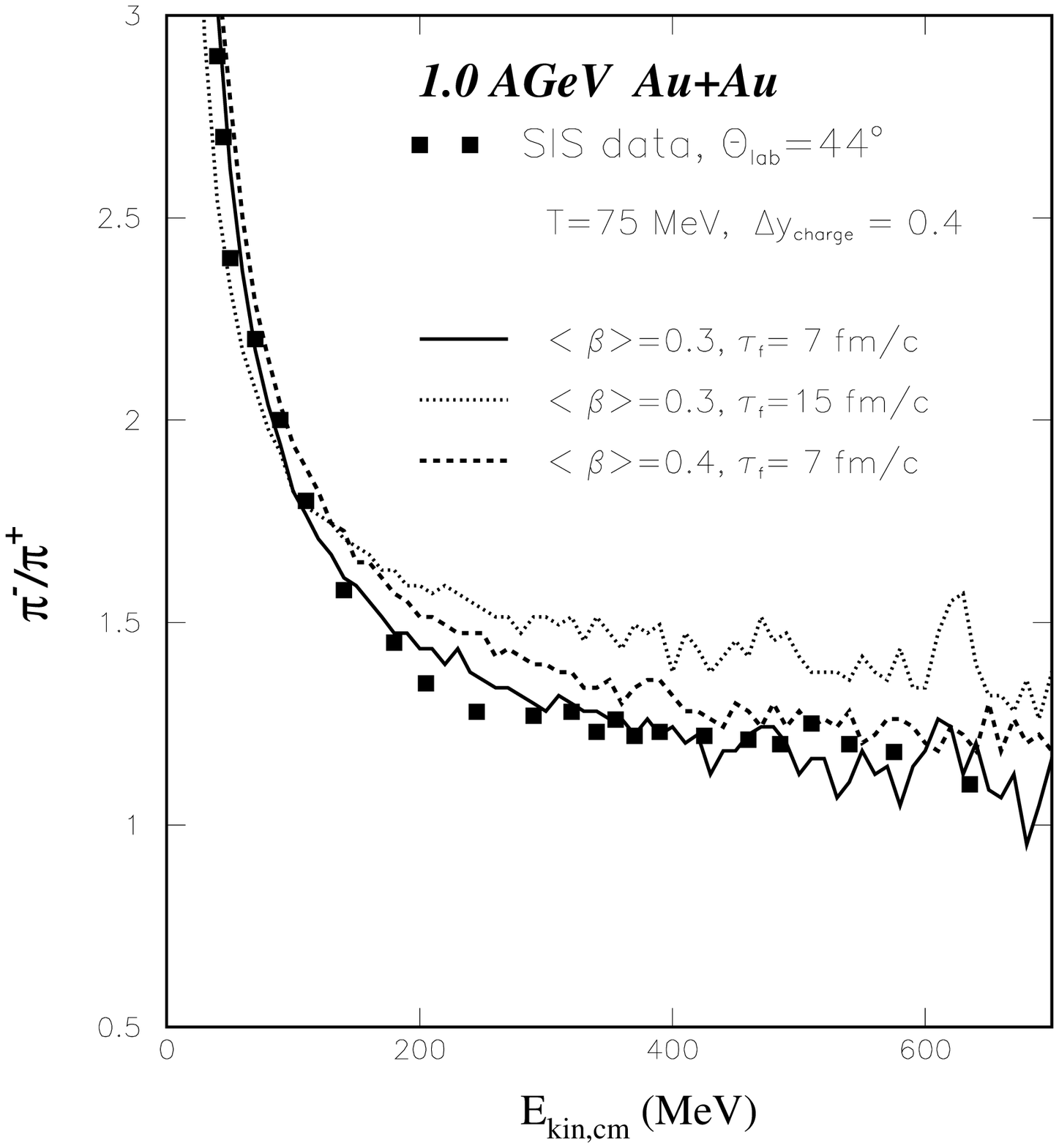,width=10cm,height=10cm,angle=0}}
\caption{
Calculations of $\pi^-/\pi^+$ ratios 
with the dynamical model 
for the reactions $Au+Au$ at 1.0 AGeV
as a function of pion kinetic energy in the center-of-mass system
compared to experiment \protect\cite{Wagner}.
Good agreement is obtained with $\tau=10 fm/c$ for a moderate
mean flow velocity $\langle \beta \rangle$ = 0.3 c or 
$\tau=7 fm/c$ for  $\langle \beta \rangle$ = 0.4 c.
\label{sisfig2} }
\end{figure}

\begin{figure}
\centerline{
\psfig{figure=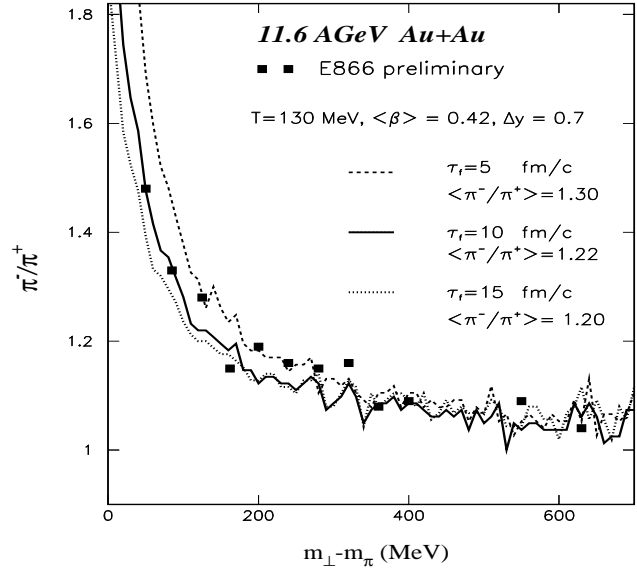,width=10cm,height=10cm,angle=0}}
\caption{
Experimental $\pi^-/\pi^+$ ratios \protect\cite{AGS}
for the reactions $Au+Au$ at 11.6 AGeV
as a function of transverse mass compared to
calculations using the dynamical
Coulomb model for different freeze-out times $\tau_f$.
\label{agsfig2} }
\end{figure}

\begin{figure}
\centerline{
\psfig{figure=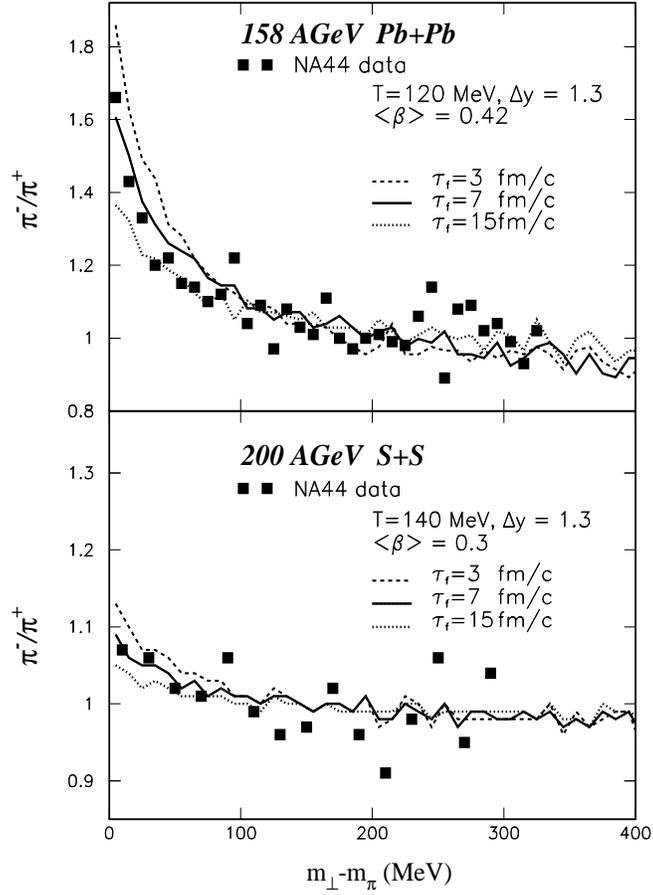,width=10cm,height=15cm,angle=0}}
\caption{
 Experimental $\pi^-/\pi^+$ ratios \protect\cite{NA44} at SPS energies
for the reactions $Pb + Pb$ and $S + S$ 
as a function of transverse mass compared to  
calculations employing the dynamical 
Coulomb model with three different values for  the freeze-out time.
\label{spsfig2} }
\end{figure}

\begin{figure}
\centerline{
\psfig{figure=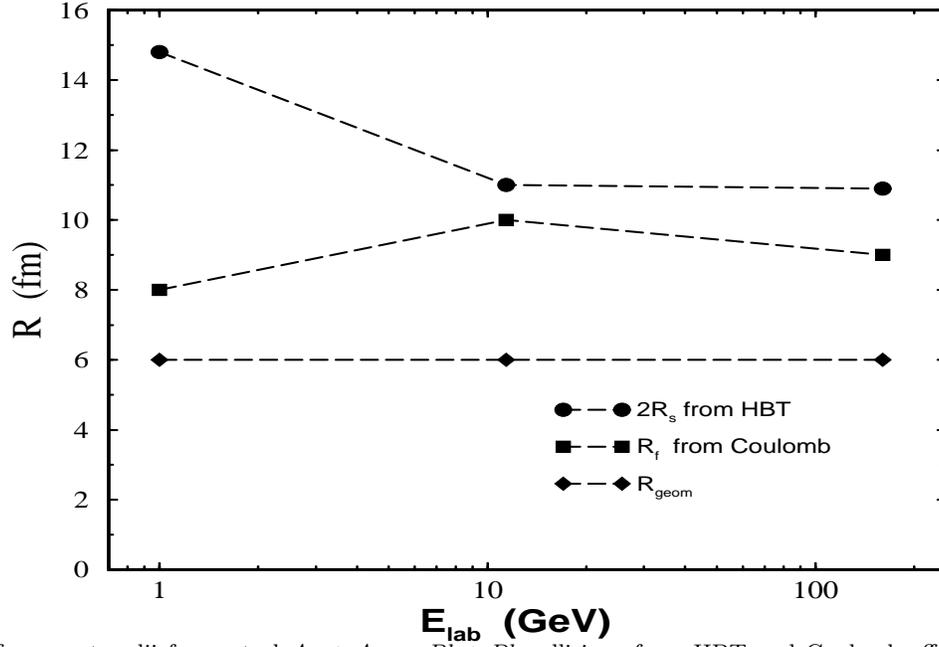,width=10cm,height=15cm,angle=-90}}
\caption{The freeze-out radii for central $Au+Au$ or $Pb+Pb$ collisions
from HBT and Coulomb effects.
The HBT data are average values of
$\pi^+\pi^+$ and $\pi^-\pi^-$ sideward HBT radii
taken from Refs. \protect\cite{SISHBT}, \protect\cite{AGSHBT}
and \protect\cite{Franz,NA49HBT,WA98}
at SIS, AGS and SPS energies respectively. 
The radii extracted from Coulomb effects
are compatible with those from HBT except at SIS energies.
The geometrical radii of the overlap zones (for $\sim14\%$ centrality)
at time of collision are smaller indicating that significant expansion 
takes place between collision and freeze-out.
\label{hbtfig} }
\end{figure}

\begin{figure}
\centerline{
\psfig{figure=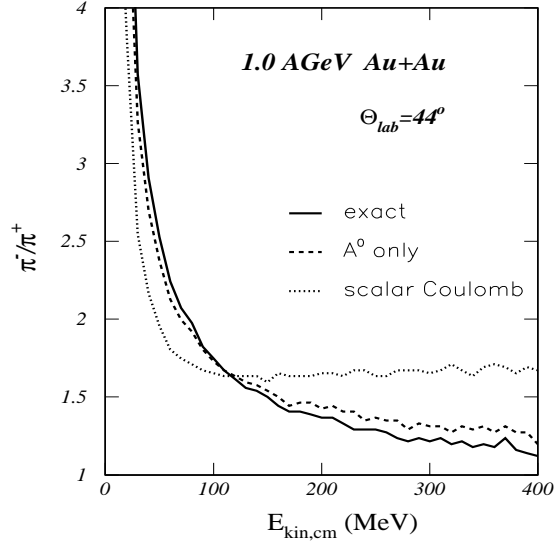,width=10cm,height=10cm,angle=0}}
\vspace{-1cm}
\caption{
  $\pi^-/\pi^+$ ratios for Au on Au collisions at 1 AGeV
as a function of the center-of mass kinetic energy
calculated with the correct retarded electromagnetic potential
(solid line) in comparison with an approximative treatment
retaining only the time component without retardation (dashed line) 
and using a simple scalar Coulomb potential (dotted line). 
\label{sisfigap} }
\end{figure}

At moderate energies where the thermal velocities of the nucleons
are not large this approximations works well.
In Fig. \ref{sisfigap} we show the effect on the $\pi^-/\pi^+$ ratio 
in the transverse direction kinetic energy
for a reaction of  gold on gold nuclei.
As an example we use the following 
parameters: $\Delta y_{ch} = 0.4$, $T= 75$  MeV, 
$\langle \beta \rangle = 0.3$, and $\tau_f =$ 7 fm/c 
which give in Fig. \ref{sisfig2} 
reasonable agreement with the data.
One recognizes that the approximate treatment (\ref{eq_approx}) is well
applicable and leads
roughly to a 10\%  underestimation of the ratio at an lab-angle of 44$^o$
which corresponds nearly to the transverse direction in the c.m. system.

\begin{figure}
\centerline{
\psfig{figure=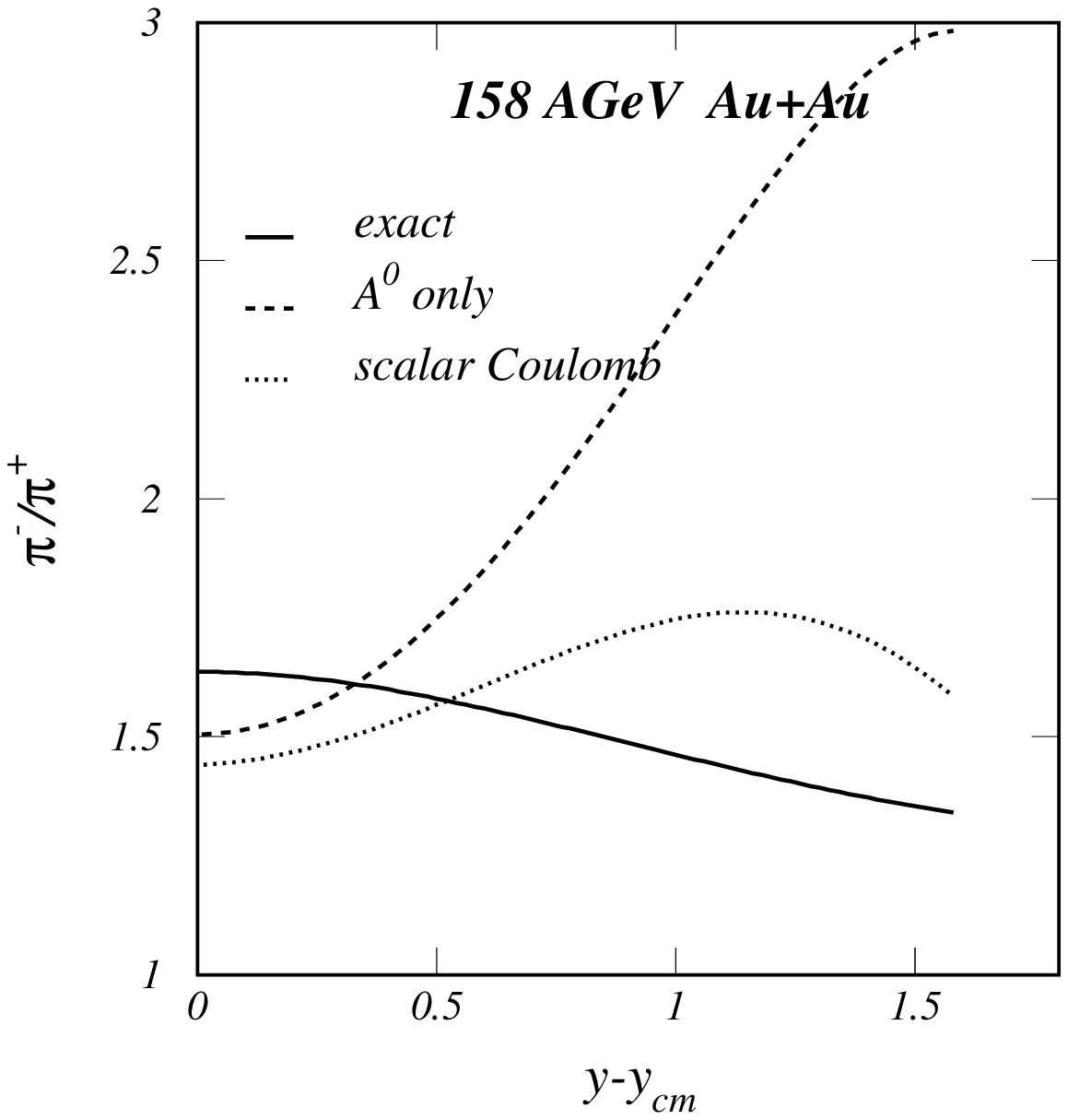,width=10cm,height=10cm,angle=0}}
\vspace{-1cm}
\caption{
  $\pi^-/\pi^+$ ratios for Pb on Pb collisions at 158 AGeV
at transverse kinetic energies  smaller than 10 MeV 
calculated with the correct potential
and two approximations (see caption of Fig. \ref{sisfigap}). 
\label{spsfigap} }
\end{figure}

Larger effects are expected at ultrarelativistic energies. 
In Fig. \ref{spsfigap}
we show the calculated charge ratio at small momentum as a function of
rapidity
for a lead on lead collisions at 158 A$\cdot$GeV. 
The solid line is calculated using Eq. (\ref{eq_motion}) and shows 
a $\pi^-/\pi^+$ ratio at zero transverse momentum which follows nicely
the rapidity distribution of the net charge as it is expected from our
considerations in section III. The approximate method shows quite a
different behavior with a maximum off-midrapidity. This is because
the potential $\phi$ is not able to cope with the translational invariance
which is hidden in the correct potential (\ref{A_mu}). Since the calculation
is carried out in the cm-system it  fails completely for larger rapidities.
In the correct treatment the action 
of the component $A^0$ is compensated by the longitudinal part of the 
magnetic field. 
The dotted curve shows the result if one treats the Coulomb potential
as a genuine scalar potential which means dropping the term $u^0$ on the
right hand side of Eq. (\ref{eq_approx}). This method is 
quite an insufficient
approximation which even is not applicable for small energies 
as can be seen in Fig. \ref{sisfigap}.


\noindent 

{\bf Figure captions}\\

\begin{thebibliography}{99}
  \bibitem{early}  W. Benenson et al., 
                   Phys. Rev. Lett. {\bf43}, 683 (1979);\\
                   K. G. Libbrecht and S.E. Koonin,
                   Phys. Rev. Lett. {\bf43}, 1581 (1979)
  \bibitem{Wagner} A. Wagner et al., preprint Th Darmstadt,  
                   IKDA 95/9 (1995) and submitted to Phys. Lett. 
  \bibitem{Pelte} D. Pelte et al.,  Z.Phys.{\bf A357} (1997) 215
  \bibitem{AGS}   L. Ahle et al., Nucl. Phys. {\bf A610} (1996) 139c  
  \bibitem{NA44}  H. B{\o}ggild et al., NA44 collaboration, 
                  Phys. Lett. {\bf B 372}, 343 (1996) 
  \bibitem{Gyulassy} M. Gyulassy and S.K. Kauffmann, Nucl. Phys. {\bf A362}
                    503 (1981)
  \bibitem{Stock} R. Stock, Phys. Rep. {\bf 135} (1986) 259.
  \bibitem{Bao}   Bao An Li,  Phys. Lett. {\bf B346} (1995) 5.
  \bibitem{Osada} T.Osada, S. Sano, M. Bijyajima and G. Wilk, Phys. Rev. 
                  {\bf C54} (1996) 54; hep-ph/9702347
  \bibitem{Barz}  H.W. Barz et al., Phys. Rev. {\bf C56} (1997) 1553
  \bibitem{Muentz}C. M\"untz et al., Z. Phys. {\bf A357} (1997) 399
  \bibitem{BB} G. Baym and P. Braun-Munzinger, {\em Nucl. Phys.} {A 610}
		(1996) 286c.
  \bibitem{Teis}  S. Teis et al., Z. Phys. {\bf A356} (1997) 421
  \bibitem{Jackson} J. D. Jackson, ``Classical Electrodynamics'', NY, Wiley
  \& Sons,1963.
  \bibitem{NA49} S.V. Afanasev, NA49 collaboration, 
                 Nucl. Phys. {A610}, 76c (1996)
  \bibitem{NA44prot} I. G. Bearden et al., 
                   NA44 collaboration, Phys. Lett. {\bf B388}, 431 (1996);
                   Nu Xu et al. (NA44), Nucl. Phys. {\bf A610}, 175c (1996)
  \bibitem{Bjorken} J.D. Bjorken, Phys. Rev. {\bf D27}, 140 (1983)
  \bibitem{NA44slopes} I.G. Bearden et al., NA44 collaboration,
                        Phys. Rev. Lett. {\bf 78}, 2080 (1997).
  \bibitem{AGSslopes} P. B. Braun-Munzinger et al., 
               Phys. Lett. {\bf B 344}, 43 (1995)
  \bibitem{SPSratio} 
       P. Jones, NA49 collaboration, Nucl. Phys. {\bf A610} (1996) 188c.  
  \bibitem{Ayala} A. Ayala and J. Kapusta, Phys. Rev. {\bf C56} (1997) 407.
  \bibitem{Herrmann} N. Herrmann et al. (FOPI collab.), 
                     Nucl. Phys. {A610}, 49c (1996).
  \bibitem{HH}    H. Heiselberg, {\em Phys. Lett.} {\bf B379} (1996) 27.
  \bibitem{Wiedemann} U.A. Wiedemann, U. Heinz, nucl-th/9611031. 
  \bibitem{Franz} I.G. Bearden et al., NA44 collaboration, 
                   Nucl. Phys. {\bf A610} (1996) 240c.
  \bibitem{fluc}  H. Heiselberg and A. Vischer, Z. Physik {\bf C}
                  (1997) in press; Phys. Lett. {\bf B} (1997) in press. 
  \bibitem{Sorge}  H. Sorge et al., Phys. Lett. {\bf B373}, 16(1996).
  \bibitem{SISHBT} D. Pelte, FOPI collaboration, proc. of Int. Conf.
                   4$\pi$ detector systems, Poiana Brasov, 1996,
                   Rumania.
  \bibitem{AGSHBT} J. Barrette et al., E877 collaboration, Nucl. Phys. 
                  {\bf A610} (1996) 227c. 
  \bibitem{Murray} M. Murray et al., NA44 collaboration, 
                   Heavy Ion Physics {\bf 4} (1996) 213.
  \bibitem{NA49HBT} K. Kadija et al., NA49 collaboration, {\bf A610} (1996) 248c.
  \bibitem{WA98}L. Rosselet et al., WA98 collaboration, {\bf A610} (1996) 256c.
  \bibitem{Kspectr} H. B\"oggild et al., NA44 collaboration, 
                      {\it Z. Phys.} {\bf C69}, 621 (1996); and preliminary
                    results presented at ``Quark Matter 1997'', Tsukuba, Japan.
  \bibitem{Koch} V. Koch, Nucl. Phys. {\bf A591}, 531c (1995).
  \bibitem{Weise} see, e.g., W. Weise, Nucl. Phys. {\bf A610} (1996) 35c.
  \bibitem{braun}  P. Braun-Munzinger et al., 
                   Phys. Lett. {\bf B 365}, 1 (1996)  
\end{thebibliography}
\end{document}